\documentclass[%
 reprint,
superscriptaddress,
 amsmath,amssymb,
prb,
]{revtex4-1}

\usepackage[utf8]{inputenc}
\usepackage[T1]{fontenc}
\usepackage{amsmath}
\usepackage{url}
\usepackage{physics}
\usepackage{graphicx}
\usepackage{dcolumn}
\usepackage{bm}
\usepackage{ulem}
 \usepackage{cancel}
 \usepackage{array,tabularx,xcolor}
\usepackage[french,english]{babel}	% police et gestion des accents?
\usepackage[squaren,Gray]{SIunits}
\usepackage[hidelinks]{hyperref}
\usepackage[all]{hypcap}

\begin{document}

\preprint{APS/123-QED}

\title{Deterministic assembly of a charged quantum dot - micropillar cavity device.}% Force line breaks with \\
%\thanks{Footnote to title of article.}

\author{P. Hilaire}
\email{present address: paulhilaire@vt.edu}
% \affiliation{C2N, Center of Nanosciences and Nanotechnology, Universit\'{e}  Paris-Sud,  Universit\'{e}  Paris-Saclay,  91120  Palaiseau,  France}
\affiliation{Université de Paris, Centre for Nanoscience and Nanotechnology (C2N), F-91120 Palaiseau, France}

\author{C. Millet}
\affiliation{Centre for Nanoscience and Nanotechnology (C2N), CNRS, Univ Paris Sud, Université Paris-Saclay, F-91120 Palaiseau, France}

\author{J.C. Loredo}
\affiliation{Centre for Nanoscience and Nanotechnology (C2N), CNRS, Univ Paris Sud, Université Paris-Saclay, F-91120 Palaiseau, France}

\author{C. Ant\'on}
\affiliation{Centre for Nanoscience and Nanotechnology (C2N), CNRS, Univ Paris Sud, Université Paris-Saclay, F-91120 Palaiseau, France}

\author{A. Harouri}
\affiliation{Centre for Nanoscience and Nanotechnology (C2N), CNRS, Univ Paris Sud, Université Paris-Saclay, F-91120 Palaiseau, France}

\author{A. Lema\^itre}
\affiliation{Centre for Nanoscience and Nanotechnology (C2N), CNRS, Univ Paris Sud, Université Paris-Saclay, F-91120 Palaiseau, France}

\author{I. Sagnes}
\affiliation{Centre for Nanoscience and Nanotechnology (C2N), CNRS, Univ Paris Sud, Université Paris-Saclay, F-91120 Palaiseau, France}

\author{N. Somaschi}
\affiliation{Quandela, 10 boulevard Thomas Gobert, 91120, Palaiseau, France}

\author{O. Krebs}
\affiliation{Centre for Nanoscience and Nanotechnology (C2N), CNRS, Univ Paris Sud, Université Paris-Saclay, F-91120 Palaiseau, France}

\author{P. Senellart}
\affiliation{Centre for Nanoscience and Nanotechnology (C2N), CNRS, Univ Paris Sud, Université Paris-Saclay, F-91120 Palaiseau, France}

\author{L. Lanco}%
\email{loic.lanco@univ-paris-diderot.fr}
% \affiliation{C2N, Center of Nanosciences and Nanotechnology, Universit\'{e}  Paris-Sud,  Universit\'{e}  Paris-Saclay,  91120  Palaiseau,  France}
\affiliation{Université de Paris, Centre for Nanoscience and Nanotechnology (C2N), F-91120 Palaiseau, France}

\date{\today}

\begin{abstract}
    Quantum-dot based spin-photon interfaces are highly sought systems to implement deterministic photon-photon gates as well as to generate photonic cluster states. This requires mastering the technological challenge of fully  controlling the coupling of a charged quantum dot to a cavity mode.  Here, we report on a set of technological and experimental developments that allows doing so. The first ingredient consists in combining the in-situ lithography technique, that allows a deterministic spatial and spectral coupling of the emitter to a pillar cavity mode, with a pre-identification of the quantum dot charge states. The second ingredient relies on the design of  an asymmetric tunneling barrier to inject the carrier in the quantum dot and  an optical control of the charge occupation probability. We show that we can ensure both a high  occupation probability of the charge in the quantum dot and an optimal coupling to the cavity mode. This is demonstrated  through second order auto-correlation measurements  and by measuring the performance of the device as a bright source of indistinguishable single photons.
\end{abstract}

\maketitle

The efficient interfacing of single photons to natural or artificial atoms is central to the development of efficient quantum light sources\cite{Varnava2008, Senellart2017}, quantum repeaters~\cite{Kimble2008, Borregaard2019}, photon-photon gates~\cite{Koshino2010, Rosenblum2011} as well as for the generation of photonic cluster states~\cite{Lindner2009, Schwartz2016}. Strong  atom-photon interactions are commonly obtained  by inserting a single atom in an optical resonator to control the spontaneous emission into a well designed optical mode~\cite{Imamoglu1999, Wilk2007}.
% The atom-photon interface can act as an efficient emitter of indistinguishable, single photons \cite{Somaschi2016, Ding2016, Wang2019}, or as an efficient single-photon receiver to use the  atom as a quantum memory to store the photonic state, and to realize various atom-photon, atom-atom or photon-photon gates~\cite{Hu2008, Koshino2010, Duan2004}.
The atom-photon interface can act as an efficient emitter of indistinguishable, single photons \cite{Somaschi2016, Ding2016, Wang2019}, but also as an efficient single-photon receiver. In the latter case, the atom can be used as a quantum memory to store the photonic state, or to realize various atom-photon, atom-atom or photon-photon gates~\cite{Hu2008, Koshino2010, Duan2004}.

Semiconductor quantum dots (QDs) are excellent artificial atoms, able to emit highly-pure single-photons and to route light at the single-photon level when inserted in cavities \cite{Rosenblum2011, Chang2014}. Exploiting the spin state of a single charge in the QD, either an electron or a hole, is essential to demonstrate deterministic photon-photon gates and generate highly sought photonic cluster states \cite{Economou2010, Pichler2017}, two central features for the scalability of optical quantum technologies.

However, the deterministic fabrication of such singly-charged cavity quantum electrodynamics (cQED)  devices is very challenging as it requires fulfilling  multiple stringent requirements at the same time. The first one is the optimal spatial and spectral coupling of a singly-charged QD state to a cavity mode. A second challenge is to controllably prepare the QD in the desired charge state, consisting of either an electron or a hole in excess. Finally, the cavity must allow the photons to be efficiently injected and collected after having interacted with the QD. In this respect, pillar microcavities allow for both efficient injection~\cite{Rakher2009, Loo2012, Hilaire2018} and collection of photons~\cite{Somaschi2016, Ding2016, Wang2019}.

Here, we report on a set of experimental and technological methods that result in a high control on the injection of a single charge in a quantum dot, which is at the same time optimally coupled to a micropillar cavity mode.  To do so, we make use of an engineered asymmetric band structure that hinders the tunneling of an optically-injected hole out of the QD.  An in-plane magnetic field photoluminescence  study of the various QD charge states allows us to identify the positively-charged  state within the pattern of various emission lines.
We define electrically connected micropillar cavities  centered on chosen quantum dots, using a cryogenic  in-situ lithography step~\cite{Dousse2008, Nowak2014}, during which we measure the positive trion energy transition and tune the cavity frequency accordingly. The quality of the fabricated charged quantum dot-cavity interfaces is  evaluated via intensity correlations measurements and by showing the excellent performance of the device as a single-photon source.

This paper is organized as follows.
In Sec.~\ref{sec_band_structure}, we present the sample structure and the technological procedure.
The scheme used to trap a single carrier in the QD is presented in Sec.~\ref{sec_control}.
We then discuss the deterministic coupling of the cavity to the trion transition in Sec.~\ref{sec_coupling} and evaluate its performances in Sec.~\ref{sec_perf}.
% The deterministic coupling of the cavity to the trion transition is discussed in Sec.~\ref{sec_coupling}, and evaluate its performances in Sec.~\ref{sec_perf}.
Finally, the properties of the devices operating as single-photon sources are discussed in Sec.~\ref{sec_source}.

\section{Sample structure and technological procedure.}
\label{sec_band_structure}

The micropillars are fabricated from a planar sample embedding a \(\lambda\)-cavity, surrounded by two distributed Bragg reflectors (GaAs and Al$_{0.9}$Ga$_{0.1}$As, with 14 and 28 pairs for the top and bottom mirrors respectively).
The reduced number of layer pairs in the top mirror allows for a higher output coupling efficiency of the photons emitted by the QD towards the top.

\begin{figure}[!ht]
    \centering
    \includegraphics[width=8cm]{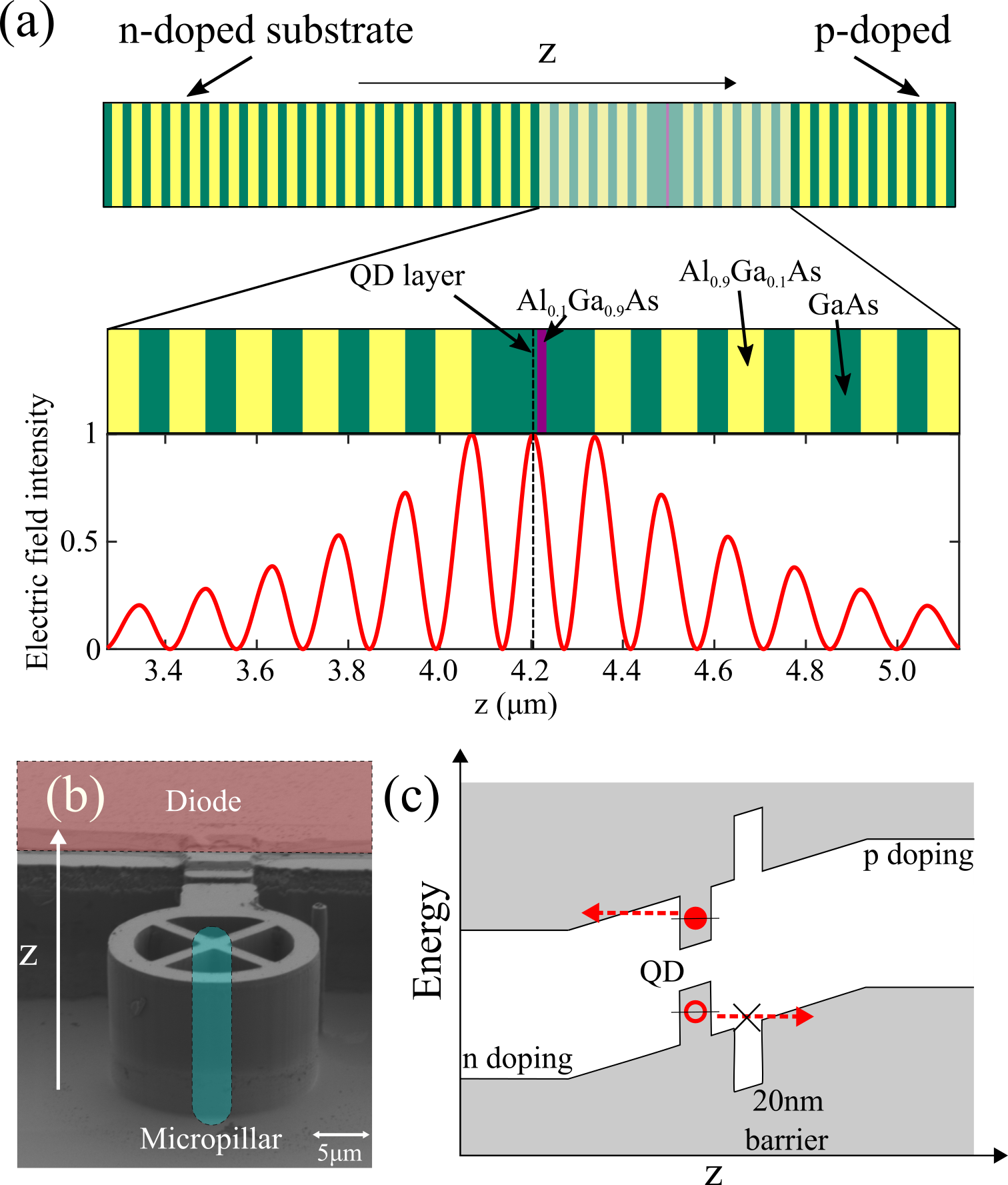}
    \caption{
    (a) Simulated electromagnetic field intensity inside the micropillar structure. The electromagnetic field is confined at the cavity layer position (green: GaAs layers,  yellow: Al$_{0.9}$Ga$_{0.1}$As layers, purple: Al$_{0.1}$Ga$_{0.9}$As tunneling barrier).
    (b) Scanning electron microscopy image of a sample.
    (c) QD + doping structure: the n-doping and p-doping region tilt the forbidden band so that fluctuating charges remain far from the QD, therefore stabilizing the electrical fluctuations. The tunneling barrier reduces the hole tunneling rate.}
    \label{fig_sample}
\end{figure}

Fig.~\ref{fig_sample}(a) displays the  electromagnetic field intensity calculated by a standard transfer matrix method as a function of the position in the structure (due to its small width, the QD layer was neglected in these simulations).
The micropillar structure  confines the electromagnetic field with a maximum intensity in its center where the quantum dot is located.

The vertical structure comprises a p-i-n junction~\cite{Prechtel2016, Rakher2009} with a doping density that gradually decreases while approaching the cavity (negative doping for the bottom mirror and positive doping for the top mirror).
Such an approach allows minimizing optical losses due to doping in the cavity layer.
The structure is connected to an electrically-contacted diode through four ridges and a circular frame as shown in Fig.~\ref{fig_sample}(b).

A well-known technique to control the charge state of QDs is via the bias voltage, which brings the Fermi energy close to a charged QD state~\cite{Drexler1994, Warburton2000}.
Here, we adopt the approach proposed in Ref.~\onlinecite{Ardelt2015}, based on an asymmetric confinement of an optically-injected electron-hole pair.
The $\lambda$-cavity consists of a GaAs layer that embeds a single InGaAs QD and a 20-nm thick tunneling barrier of Al$_{0.1}$Ga$_{0.9}$As, positioned 10nm above the QD layer.
The energy levels of the QD and its near environment are schematized in Fig.~\ref{fig_sample}(c), which sketches the energy bands as a function of the vertical position in the cavity layer.
The doped regions tilt the bands, allowing the tunneling of carriers out of the quantum dot. Yet the Al$_{0.1}$Ga$_{0.9}$As barrier, being placed above the QD layer, is used to strongly increase the tunneling time for a hole trapped in the quantum dot, typically by three orders of magnitude~\cite{Ardelt2015}, while keeping the electron tunneling time unchanged.
The resulting difference in the electron-hole tunneling time allows the trapping of a single hole when optically creating  an electron-hole pair.

The micropillar cavities are fabricated using the in-situ lithography technique to center the  chosen QDs within the pillar mode~\cite{Dousse2008, Nowak2014}. This is obtained by measuring the QD position with nanometer-scale accuracy through photoluminescence mapping and by defining a connected pillar cavity  centered on the QD  during the same step.  The cavity diameter is adjusted so that the trion transition is matched to the cavity mode. It is thus essential to be able to identify the QD emitting charge state (neutral exciton, charged exciton, etc) during the in-situ lithography procedure. To do so, we detail the experimental conditions used here to optically trap a hole in Sec.~\ref{sec_control} and discuss how we  identify the lines through magnetic field spectroscopy in Sec.~\ref{sec_id}.

\section{Optical injection of a single-hole.}
\label{sec_control}

\begin{figure}[!ht]
	\centering
	\includegraphics[width=7cm]{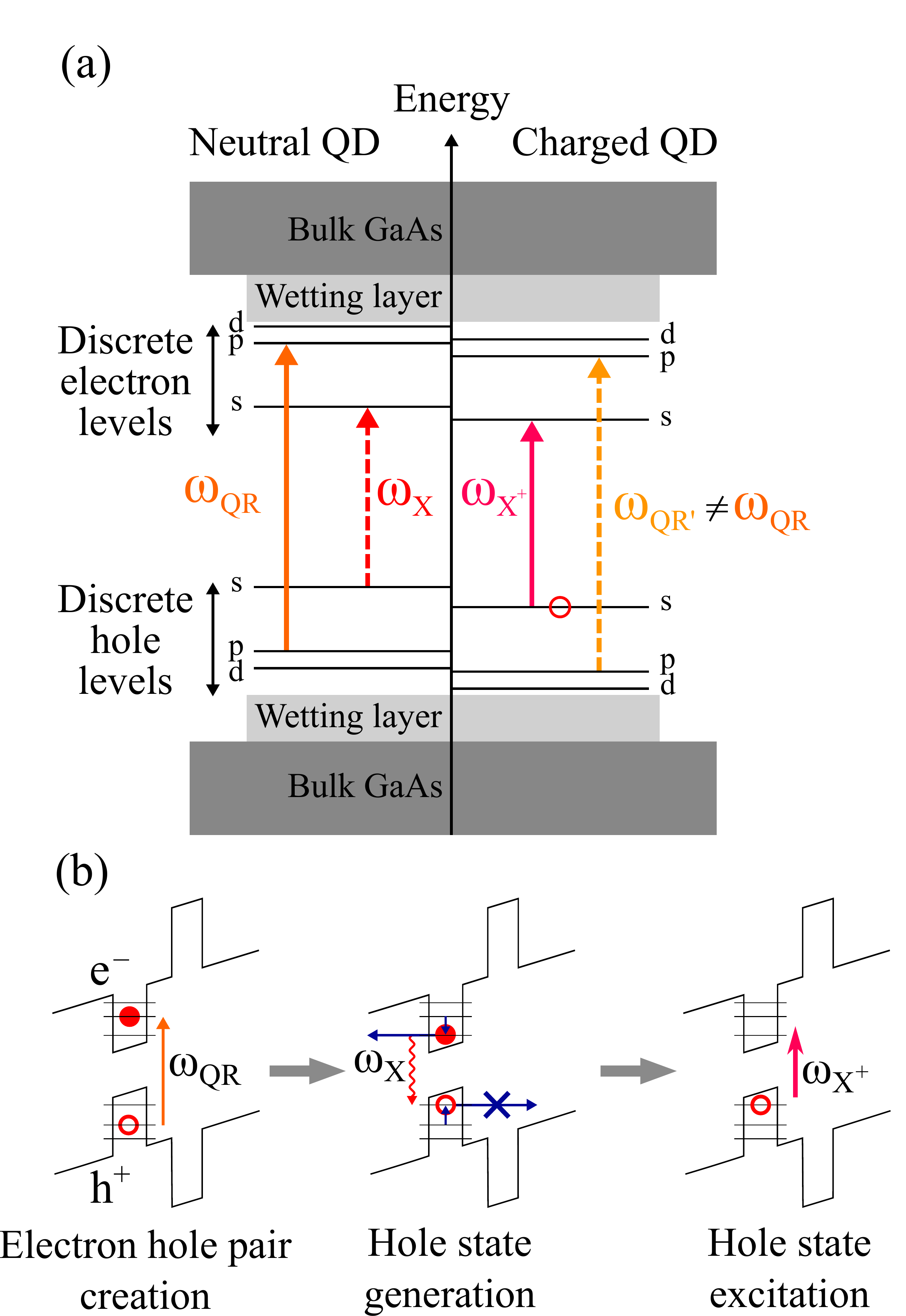}
	\caption{
    (a) Sketch comparing the energy levels of a neutral QD and a singly-charged QD. The resonant trion transition energy $\omega_{X^+}$ is different from the exciton transition energy $\omega_X$ ($\omega_{X^+} \neq \omega_X$) and the p-shell transition energies of a charged QD are also different from a neutral QD: $\omega_{QR'} \neq \omega_{QR}$.
	(b) Hole trapping scheme using quasi-resonant excitation.
	Left panel: the QD stable state is the crystal ground state. A quasi-resonant laser creates an electron-hole pair by exciting a p-shell transition ($\omega_{QR}$).
    Middle panel: If the electron tunnels out the QD before radiative recombination with the hole, it generates of a single hole QD state.
	% Middle panel: The generation/recombination process of electron-hole pairs keeps going until the electron tunnels out while the hole is blocked by the barrier, thus generating a positive singly-charged QD state.
	Right panel: the QD can then be resonantly excited by a laser with energy $\omega_{X^+}$ corresponding to the trion transition.}
	\label{fig_nr_scheme}
\end{figure}

We describe here the optical method used to  inject a hole in the quantum dot. The stable QD state is the crystal ground state, here denoted as neutral QD, whose energy levels are represented in the left part of Fig.~\ref{fig_nr_scheme} (a).

An electron-hole pair can be generated by resonantly pumping the exciton ($\omega_X$) or other discrete transitions (such as "p-shell" transitions or LO-phonon-assisted transitions) here denoted as quasi-resonant (QR) transitions with energy $\omega_{QR} >\omega_X$~\cite{Pooley2012, Reindl2019}. Here, this is done using a CW laser at a wavelength around $\lambda_{QR} = 901 \nano\meter$ as depicted in the left panel of Fig.~\ref{fig_nr_scheme}(b).

When an exciton is quasi-resonantly created in the QD, the electron and the hole excess energy non-radiatively decays (in typically less than $100\pico\second$ \cite{Sosnowski1998})  and the system reaches the lower exciton level, as shown in the middle panel of Fig.~\ref{fig_nr_scheme}(b).
If the exciton radiatively recombines by emitting a photon, the QD then returns to its ground state.
This generation-recombination cycle keeps going until an optically-created electron eventually tunnels out towards the n-doped electrical contact before the radiative recombination occurs.
Due to the Al$_{0.1}$Ga$_{0.9}$As barrier, the remaining hole is confined in the QD for a longer time, during which the QD is in the desired positively-charged state.

The addition of a hole in the QD induces a strong modification of the electric environment by Coulomb interaction, modifying all the discrete energy levels of the QD, as illustrated in the right-hand side of Fig.~\ref{fig_nr_scheme}(a).
The positively-charged QD  thus presents a different energy structure than the neutral one, a feature that applies to LO-phonon assisted and p-shell optical transitions as well.
Thus, the incoming frequency $\omega_{QR}$ does not induce any optical excitation when a hole is confined. This allows trapping only a single hole in the quantum dot as shown later on.

\section{Charge state identification.}
\label{sec_id}

\begin{figure*}[!ht]
	\centering
	\includegraphics[width=17cm]{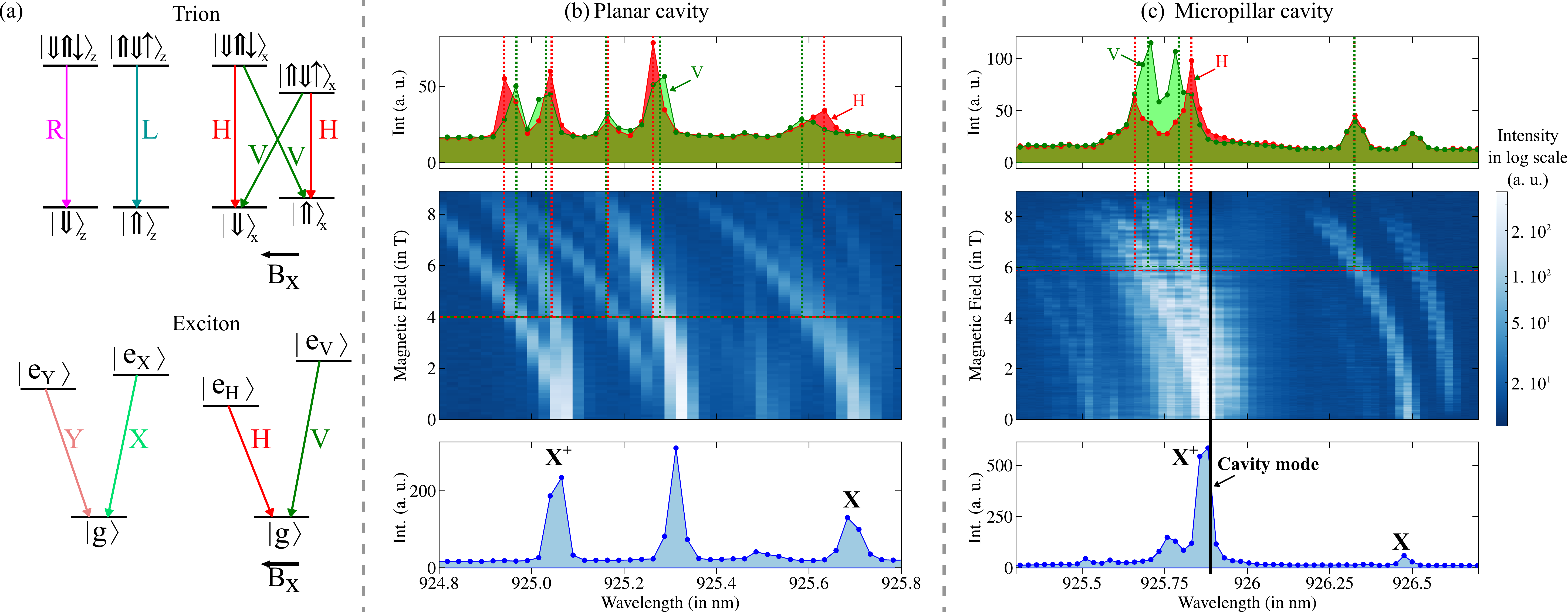}
	\caption{
    (a) Optical selection rules of a trion (top) and an exciton (bottom), with (right) and without (left) external transverse magnetic field.
	(b, c) QD photoluminescence under an $850$-nm non-resonant laser in (b) a planar cavity sample before etching and (c) after the pillar etching
   ((b) and (c) represent photoluminescence from different quantum dots.).
    Bottom panel: without any applied external magnetic field and polarization selection. Middle panel: under varying in-plane magnetic field. Top panel: with strong in-plane magnetic field ((b) $B_x = 4 \tesla$ and (c) $B_x = 6 \tesla$) measured with  horizontal (red) and vertical (green) polarizations. (In the middle panel of (c), a waveplate is also rotated before the spectrometer, during the magnetic field scan, to obtain polarization resolution.)}
	\label{fig_mag_field_th}
\end{figure*}

In order to define the pillar cavity diameter to match the trion transition of a positively-charged QD, we have developed spectroscopy tools to identify the QD energy levels in the experimental conditions of the in-situ lithography - i.e. non-resonant excitation around 850 nm. With above-band excitation, the emission spectrum of the QD presents emission lines corresponding to both the neutral and charged energy levels. To be able to distinguish these lines, we have performed a systematic  study of the QD emission under an in-plane magnetic field.

The trion transitions with and without magnetic field are sketched respectively in the top-right and top-left panels of Fig.~\ref{fig_mag_field_th}(a). At zero external magnetic field (top-left panel of Fig.~\ref{fig_mag_field_th}(a)), a trion radiatively decays into the hole state by emitting a circularly-polarized photon with either a right-handed or a left-handed helicity. There are therefore two trion transitions that are energy degenerate.
When an in-plane magnetic field is applied, it induces a Zeeman splitting between the hole states and the trion states, and it also modifies the system eigenstates and thus the optical selection rules, as illustrated in the top-right panel of Fig.~\ref{fig_mag_field_th}(a).
For a high in-plane magnetic field ($B_x>1 \tesla$ typically), a trion decays into a hole by emitting a single photon with 4 combinations of linear polarizations and energies.

The neutral exciton transitions with and without magnetic field are sketched respectively in the bottom-right and bottom-left panels of Fig.~\ref{fig_mag_field_th}(a). At zero external magnetic field, the two exciton states $\ket{e_X}$ and $\ket{e_Y}$ radiatively decay into the same ground state $\ket{g}$ (the neutral state) by emitting a photon with linear polarization X and Y, respectively. An in-plane magnetic field increases the energy splitting between the two excited states and modifies the polarization of the emitted photons to either horizontal or vertical. Note that even in the case of a strong in-plane magnetic field, an exciton decays by emitting a single photon with only two combinations of linear polarizations and energies and thus, can be discriminated from the trion transition.

The bottom panel of Fig.~\ref{fig_mag_field_th}(b) shows a typical photoluminescence spectrum obtained on a QD in a planar cavity sample, under non resonant excitation ($\lambda_{NR} = 850\nano\meter$) and without any applied magnetic field. There are three dominant transitions observed, at $925.1\nano\meter$, $925.3\nano\meter$ and $925.7\nano\meter$; the lower intensity transition (at $925.5\nano\meter$) is related to another QD spatially close to the one under study.

When an in-plane magnetic field is applied, the QD transitions split as seen in the middle panel of Fig~\ref{fig_mag_field_th}(b), which displays the evolution of the photoluminescence spectrum under an increasing in-plane magnetic field intensity. At $B_x = 4\tesla$ and above, all the observed peaks are split into two transitions. At high magnetic field, all the observed QD transitions are also blue-shifted, due to the diamagnetic shift~\cite{Bayer1998}.
The trion transition splits into four different transitions; however, the limited spectrometer resolution impedes the visualization of such effect which is revealed by polarization analysis instead.

The top panel of Fig.~\ref{fig_mag_field_th}(b) displays the polarization-resolved photoluminescence for the same QD at $B_x=4$T. The red (green) curve corresponds to the horizontally-polarized (vertically-polarized) photoluminescence. The horizontal (vertical) polarization peak centers are highlighted by red (green) vertical lines.

Let us first consider the higher wavelength transition ($925.7\nano\meter$ at $B_x = 0 \tesla$) from the bottom panel of Fig.~\ref{fig_mag_field_th}(b). The magnetic field scan and the polarization analysis show that it is split in two transitions with orthogonal linear polarizations. This is the expected behavior for an exciton~\cite{Bayer2002}.

As can be seen in the middle and top panels of Fig.~\ref{fig_mag_field_th}(b), the lower wavelength transition ($925.1\nano\meter$ at $B_x = 0 \tesla$) analysis shows that it is split in four transitions with different energies and polarizations: the highest and lowest have the same horizontal polarizations and are orthogonal to the two intermediate vertically-polarized transitions.
This state is therefore identified as a trion, as this is the behavior expected from the polarization selection rules of Fig.~\ref{fig_mag_field_th}(a).
% \blue{Commenter que le splitting n’est pas le meme pour la raie haute et basse, toujours en accord avec la theorie et que c’est signature de:} \blue{ Je ne suis pas sûr qu'on puisse dire grand chose de cela... Si c'était un électron, on aurait certainement le même résultat. C'est la signature que l'état fondamental et l'état excité ont des facteurs de Landé différents, donc que potentiellement il s'agit d'un électron et d'un trou (+ éventuellement un état singlet qui ne se splitte pas). Donc on ne peut pas déduire si c'est un trion positif ou négatif. Je ne suis pas sûr que ça apporte quelque chose à la discussion, de dire que les facteurs de Landé sont différents pour l'état excité et l'état fondamental. S'il est question du fait que les transitions H et V "hautes" et "basses" n'ont pas le même splitting entre elles. Je pense que c'était plus un problème de résolution du spectromètre qui donne cet effet...}

We note that the center transition ($925.3\nano\meter$ at $B_x = 0 \tesla$) analysis is also split in four transitions with similar polarizations as the QD trion. However, there is a clear asymmetry in the photoluminescence intensity for these four transitions. This feature may be explained by the $X^{2+}$ states, corresponding to a QD charged with two holes~\cite{Ediger2007, Ediger2007b}.

\section{Positive trion-optical mode coupling }
\label{sec_coupling}

We have conducted the above emission analysis on a large number of QDs on the planar cavity sample used to fabricate the pillars. Such a systematic study allows us to identify the various QD transitions in a fingerprint-type analysis~\cite{Mlinar2009}. During the in-situ lithography step, while no magnetic field is applied, we can thus precisely identify the lines corresponding to the charged-QD  and tune the pillar cavity mode energy to the right transition. After the in-situ lithography procedure, and the following etching and electrical contacting steps~\cite{Dousse2008, Nowak2014}, we return to magneto-optics measurements to demonstrate the coupling of the positively charged trion transition to the cavity mode.

Fig.~\ref{fig_mag_field_th}(c) presents the emission spectrum of a QD inserted in a micropillar under above-band excitation. % is then compared to the cavity mode energy to verify that they are indeed coupled: the result is displayed in .
%in a QD-micropillar device
   % \blue{rajouter cette phrase si on compare un spectre avant et après}\red{ The lines can be mapped to the spectrum before etching, considering that the energy levels of the QD shift typically by $500 \pico\meter$-shift when the strain is released through the etching process. This additional energy shift is anticipated during the in-situ lithograhy step to define the proper pillar diameter the etching process}. \green{That is not the same QD}
%After the realization of electrically-contacted QD-micropillar devices, the trion-cavity spectral matching is evaluated by repeating the procedure of trion identification under an in-plane magnetic field.
The bottom panel represents the photoluminescence at $B_x =0 \tesla$, displaying discrete peaks: the most intense peak corresponds to the QD transition that is coupled to the fundamental cavity mode.
% In the data plotted in the middle panel of Fig.~\ref{fig_mag_field_th} (c), the information on the polarization and the magnetic field dependence are acquired at the same time: a set of half waveplate and polarizer is positioned before the spectrometer and is rotated with constant speed which provides the polarization information of the QD photoluminescence, throughout the magnetic field scan.
% It
The middle panel of Fig.~\ref{fig_mag_field_th}(c) shows the evolution of the photoluminescence when increasing the magnetic field.
The top panel of Fig.~\ref{fig_mag_field_th}(c) represents the photoluminescence collected for an external transverse magnetic field of approximately $6 \tesla$ in H and in V polarizations.
The QD transition that is in resonance with the fundamental cavity mode is characterized by four Zeeman-split transitions as expected for a positive trion transition.
% \blue{Further analysis shows an acceleration of spontaneous emission of $3 \pm 0.5$, consistent with the Purcell effect expected for the pillar diameter considered here for a QD centered in the pillar center.}
% The trion transition, characterized by the four Zeeman-split transitions, is identified at 925.9nm (at $B_x = 0 \tesla$), which is indeed in resonance with the cavity mode.

\section{Performances of the hole trapping}
\label{sec_perf}

In the following, the hole trapping scheme performances, namely the single-hole occupation probability and the single-hole trapping time, are evaluated by observing the resonance fluorescence signal on multiple QD-cavity devices. In addition to the quasi-resonant laser used for the hole trapping, we use a second laser in resonance with the trion transition, i.e. at the energy $\omega_{X^+}$ ($\neq \omega_X$) with typical power $P_{X^+} \approx 0.1 - 3\nano\watt$, to measure the resonance fluorecence of the charge QD.
% In the following, we use an excitation  in resonance with the trion transition, i.e. at the energy $\omega_{X^+}$ ($\neq \omega_X$) with typical power $P_{X^+} \approx 0.1 - 3\nano\watt$, to measure the resonant fluorescence of the charged quantum dot as schematized in the right panel of Fig.~\ref{fig_nr_scheme}(b).

To suppress the resonant laser and only collect the emitted single photons, we use a cross-polarized setup where the trion transition is resonantly excited with a horizontally-polarized resonant laser. The resonance fluorescence is collected through a vertical polarizer that filters out the resonant laser (the quasi-resonant laser is spectrally filtered).
The $15$-ps pulse laser in resonance with the trion transition induces Rabi oscillations between the hole and the trion state \cite{Kamada2001, He2013, Giesz2016} depending on the pulse area $P_{X^+}$.
Note that single photons are emitted only if the hole is trapped in the QD. Therefore, the detection of a single photon ensures that a single hole is confined in the QD.

\begin{figure}[!ht]
	\centering
	\includegraphics[width=8cm]{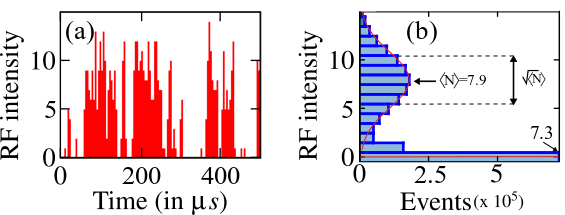}
	\caption{
    (a) Time trace recording the number of single-photons detected per time bin $\Delta t=4 \micro\second$ (with $P_{QR}=50\micro\watt$ and $P_{X^+}$ corresponding to $\pi$-pulse).
	(b) Histogram of the number of time bins corresponding to a given number of detected photons, during an acquisition time of approximately $10 \second$. It represents the intensity distribution of the single-photon source and displays the addition of a high zero detected photon probability and a gaussian distribution (with $\langle N \rangle =7.9$ photons /$\Delta t$ and a width of $\sigma = 2.8 (=\sqrt{\langle N \rangle})$).
	}
	\label{fig_fluctuations_signal}
\end{figure}

Fig.~\ref{fig_fluctuations_signal}(a) represents the time trace of the cross-polarized resonance fluorescence intensity emitted by one QD-pillar device, observed with a quasi-resonant power of $P_{QR} = 50\micro\watt$, a $\pi$-pulse resonant excitation and a time bin of $\Delta t = 4 \micro\second$. It evidences a clear blinking of the QD occupation levels between charged and neutral states. This is further shown in the histogram of Fig.~\ref{fig_fluctuations_signal}(b) representing the distribution of the number of detected events per time bin $\Delta t$.
The histogram shows the presence of two states for the QD, a bright and dark state, the latter corresponding to any case where the QD does not trap a single hole, leading to the absence of detected photons and thus $N=0$ photon per time bin. The signal corresponding to the bright state shows a typical Poisson distribution, centered on $\langle N \rangle=7.9$ photons per time bin, with a width given by the square root of the average value.

\begin{figure}[!ht]
	\centering
	\includegraphics[width=7cm]{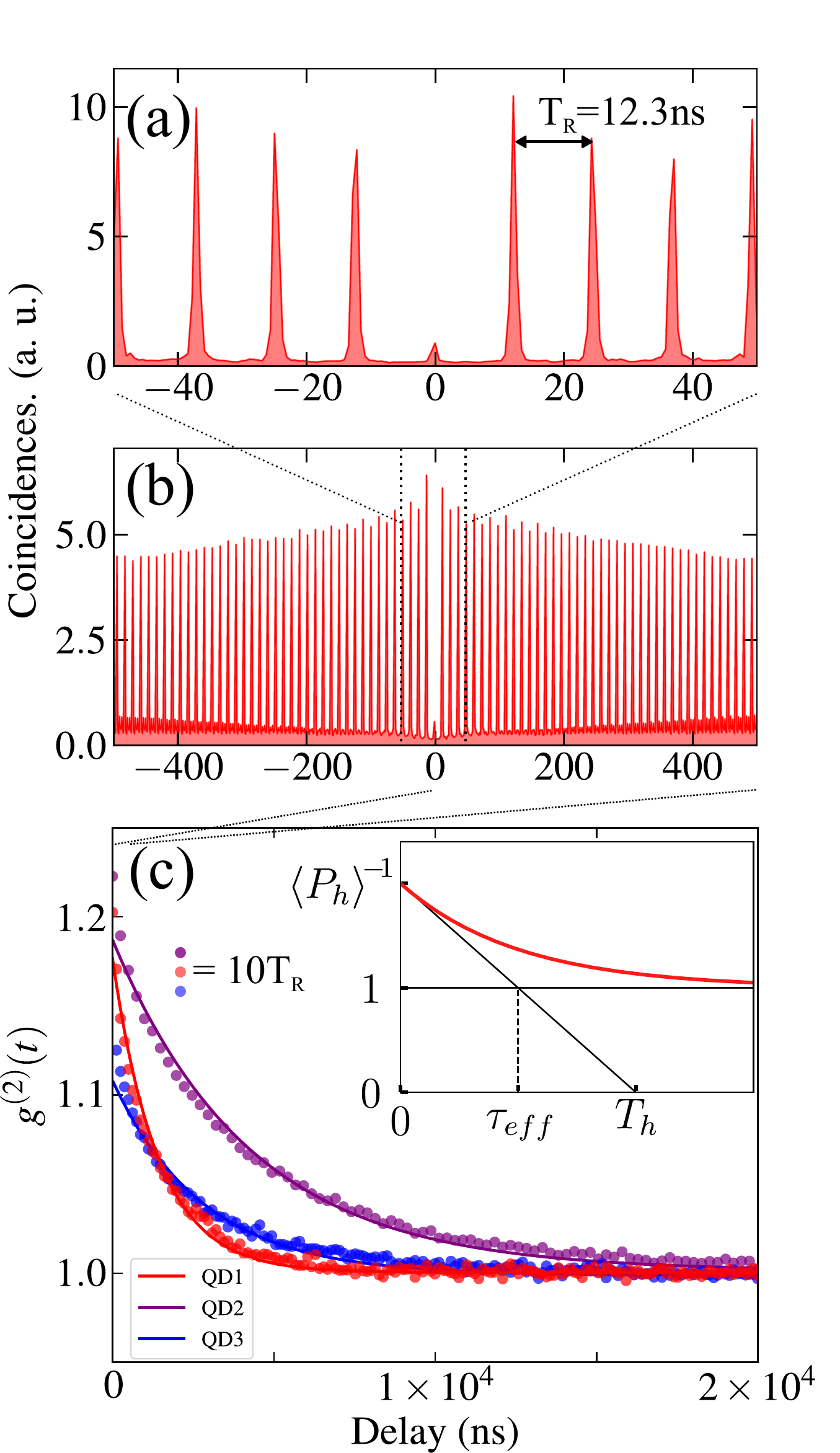}
	\caption{
    (a,b) Coincidence measurement observed at (a) short  and (b) intermediate timescales.
	(c) Intensity auto-correlations $g^{(2)}(t)$ of the same set of data (QD1) and for two other devices (QD2 and QD3), for even longer timescale and with time bin $\Delta t = 10 T_R$. The decay is fitted by an exponential curve (in red). The inset illustrates how the hole occupation probability $\langle P_h \rangle$ and the hole trapping time $T_h$ can be deduced from these correlation measurements.
	}
	\label{fig_correlations}
\end{figure}

From this histogram, comparing the area of the bright and dark state events, we can deduce a hole occupancy of around $60 \%$. Note that  there is another signal at $N = 1$ photon per time bin which prevents an accurate determination of the hole occupation probability using this method.
We identify this noise to a background signal arising from the imperfect laser rejection and the detector dark counts.
To precisely determine the hole occupation probability together with the hole trapping time, we now turn to second-order correlations of the cross-polarized resonance fluorescence as pioneered by Ref.~\onlinecite{Piketka2013}.
A typical auto-correlation measurement is displayed in Fig.~\ref{fig_correlations}(a), where the two photon coincidence histogram is plotted for different detection delays.
The peaks are separated by a delay $T_R = 1/f$, where $f = 82\mega\hertz$ is the laser repetition rate.
In Fig.~\ref{fig_correlations}(a), the continuous background between two peaks is due partly to the detector dark counts, but mainly to imperfect filtering of the quasi-resonant laser.

Fig.~\ref{fig_correlations}(b) shows the same set of data for longer delays, at which we can observe that the envelope of the peaks is decaying.
Fig.~\ref{fig_correlations}(c) shows the intensity auto-correlations $g^{(2)}(t)$ at even longer times (integrated with a time bin $\Delta t = 10 T_R$ ) with the same set of data (QD1 in red) and for two other devices (QD2 and QD3 respectively in purple and blue).
Here, the large binning blurs out the peaks which were visible in Fig.~\ref{fig_correlations}(a) and (b), as well as the zero-delay antibunching.
The correlation measurements evidence an exponential decay for the three devices, originating from the on/off fluctuations of the hole state in the QD. This effect is used to extract the hole confinement characteristics, as is shown in the following.

The auto-correlation function $g^{(2)}(t)$ can be interpreted as the probability to detect a photon in a detector (denoted APD1) at time $t$, conditioned to the detection at time $t=0$ of a photon in the other detector (denoted APD0) normalized by the uncorrelated probability of detecting a photon at any time:
\begin{equation}
	 g^{(2)}(t) = \frac{P(\mathrm{APD1}, t| \mathrm{APD0}, 0)}{P(\mathrm{APD1})}
\end{equation}
Immediately after a photon detection in APD0, the probability that a hole is trapped in the QD is $P_h(0)=1$ and the detection of a photon in APD1 is thus more probable shortly after this first detection.

To interpret the dynamics of the charge state, we develop a model using two states: "0" (the crystal ground state, with zero excess charge) and "h" (the hole state) \cite{Piketka2013}. The CW quasi-resonant laser transfers the quantum dot state from "0" to "h" at a rate $\gamma$ (which directly depends on the quasi-resonant power $P_{QR}$, used to populate the single hole state). Conversely, the hole can tunnel out from the quantum dot with a tunneling time $T_h$. The occupation probability of the empty state and the charged states are denoted $P_0(t)$ and $P_h(t)$ respectively, with $P_0(t)+P_h(t)=1$.
Their rate equations are:
\begin{equation}
\begin{aligned}
\frac{d P_h(t)}{dt} = \gamma P_0(t) - \frac{1}{T_h} P_h(t) \\
\frac{d P_0(t)}{dt} = - \gamma P_0(t) + \frac{1}{T_h} P_h(t) \\
\end{aligned}
\end{equation}

We now assume that at time $t=0$, a single photon emitted by the QD is detected. A hole is thus trapped inside the quantum dot with a probability $P_h(0) = 1$ and the hole occupation probability evolution is given by:
\begin{equation}
P_h(t) = (1 - \langle P_h \rangle) e^{-\frac{t}{\tau_{eff}}} + \langle P_h \rangle
\end{equation}
with $\langle P_h \rangle = P_h(\infty) =  \gamma / (\gamma + \frac{1}{T_h})$, the average hole occupation probability, and $\tau_{eff} = \left(\gamma + \frac{1}{T_h} \right)^{-1}$, the effective time characterizing the charge dynamics.

In this model, the enveloppe of the auto-correlation function is given by:
\begin{equation}
	 g^{(2)}(t) = \frac{P_h(t)}{\langle P_h \rangle} = \left(\frac{1}{\langle P_h \rangle} - 1\right)e^{-t /\tau_{eff}} + 1
	 \label{eq:g2_th}
\end{equation}

Therefore, it is possible to extract the hole occupation probability and the hole tunneling time through these auto-correlation measurements, as illustrated in the inset of Fig.~\ref{fig_correlations}(c):
the enveloppe of the $g^{(2)}(t)$ varies from $1/\langle P_h \rangle$ at zero delay to 1 in a characteristic time $\tau_{eff}$.
In addition, its tangent at zero-delay crosses the x-axis at $t=T_h$.

\begin{figure*}[!ht]
	\centering
	\includegraphics[width=18cm]{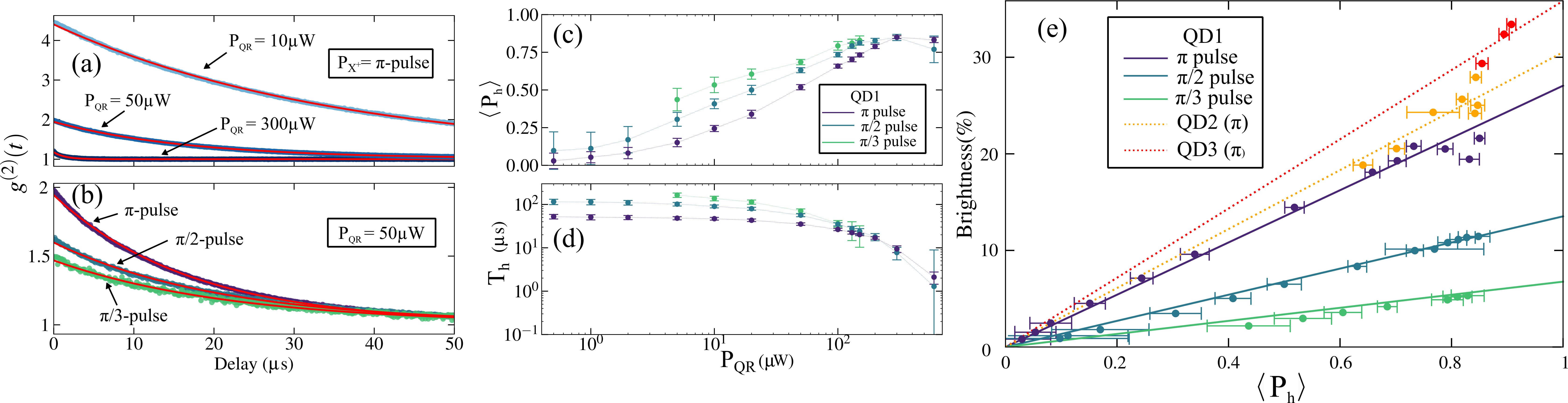}
	\caption{
    (a, b) Auto-correlation measurements for (a) varying non resonant powers $P_{QR}$ (with fixed $P_{X^+}$ corresponding to $\pi$-pulse) and (b) varying resonant power $P_{X^+}$ (with fixed $P_{QR} = 50 \micro\watt$). Red curves are exponential decay fits.
	(c,d) Extracted (c) hole occupation probability $\langle P_h \rangle$  and (d) trapping time $T_h$ as a function of the non resonant $P_{QR}$ and resonant $P_{X^+}$ powers.
	(e) Polarized brightness dependence on the hole occupation probability. Points are experimental data that are fitted with solid linear lines. The brightness measurement error bars are not represented on this panel but are typically of $\pm 5\%$.
    }
	\label{fig_occ_time}
\end{figure*}

We include the effect of small background noise and  obtain the real auto-correlation function of the quantum dot light source $g^{(2)}(t)$ deduced from the experimental one $g_{exp}^{(2)}(t)$. Let $P_{QD}$ be the probability that a photon detected is originated from the quantum dot and $1 - P_{QD}$, the probability that it is originated from bad laser filtering or dark counts. The relation between $g_{exp}^{(2)}(t)$ and $g^{(2)}(t)$ is given by:
\begin{equation}
g^{(2)}(t) = \frac{g_{exp}^{(2)}(t) - 2(1-P_{QD}) + (1-P_{QD})^2}{P_{QD}^2}
\end{equation}
In Fig.~\ref{fig_correlations}(c) and~\ref{fig_occ_time}(a,b), the corrected $g^{(2)}(t)$ is displayed, from which the occupation probability
and the hole trapping time can be directly extracted using the fit with Eq.~\ref{eq:g2_th}.

The dependence of the hole trapping time and hole occupancy  is now studied as a function of the quasi-resonant ($P_{QR}$) and resonant ($P_{X^+}$) laser powers.
Fig.~\ref{fig_occ_time}(a) shows the dependence of correlation measurements with respect to $P_{QR}$, with $P_{X^+}$ fixed to $\pi$-pulse.
Similarly, Fig.~\ref{fig_occ_time}(b) shows the dependence of correlation measurements with $P_{X^+}$, with  $P_{QR}$ fixed to $50 \micro\watt$. These two figures show that the correlation timescales and the zero-delay value both depend on the resonant and quasi-resonant laser powers.

For all $P_{QR}$ and $P_{X^+}$, the resulting correlations are fitted with an exponential decay, to extract the hole occupation probability $\langle P_h \rangle$ and the hole trapping time $T_h$, that are displayed in Fig.~\ref{fig_occ_time}(c) and (d), respectively.
As expected, $\langle P_h \rangle$ is increasing with $P_{QR}$: it reaches $\langle P_h \rangle = 85 \pm 1\%$ for QD1. The hole trapping time is higher than $20 \micro\second$ except for $P_{QR}> 100 \micro\watt$.
This shows that $T_h$ is always higher than the typical hole spin lifetime at zero magnetic field (which is generally around $1 \micro\second)$~\cite{Dahbashi2014}.

We note that $T_h$ and $\langle P_h \rangle$ should not vary with the resonant power according to the simple model considering only the zero and one hole states.
In addition, the hole trapping time $T_h$ should be constant, equal to the hole tunneling time.
The observed deviation from our model can be explained if we consider the possibility of generating a two-hole state: due to the pumping by the two lasers, it is possible to generate a second electron-hole pair when a single hole is already present in the QD and, if the electron tunnels out of the quantum dot, to obtain a two-hole state. This limits the hole occupation probability and reduces the hole trapping time in the strong pumping regime.

Finally, Fig.~\ref{fig_occ_time}(e) shows the evolution of the polarized brightness $B_p$ as a function of the measured hole occupation probability.  $B_p$ is defined as the probability  per excitation pulse to detect a polarized single photon after the first lens~\cite{Senellart2017}. It is measured by dividing the measured count rate $C$, by the laser repetition rate $f$, the setup transmission $T$ (evaluated by independant component transmission measurements), and the detector efficiency $\eta_{det}$:
\begin{equation}
    B_p = \frac{ C}{f T \eta_{det}}
\end{equation}

As expected, the measured brightness is  proportional to the hole occupation probability : the measured $B_p$ corresponding to $\pi$, $\pi/2$, $\pi/3$ pulses are fitted altogether by 3 linear functions with relative slopes $s_{\pi}=26.2\%$, $s_{\pi/2}=s_{\pi} \cos^2{(\pi/2)}$ and $s_{\pi/3}=s_{\pi} \cos^2{(\pi/3)}$
(the $\cos^2$ takes into account the partial population inversion to the trion state for $\pi/2$ and $\pi/3$ pulses).

% Leads to intracavity photons escaping the cavity from the top with a probability $\eta_{top} = 85 \pm 5 \%$.
The probability that an intracavity photon escapes the cavity through the top mirror is called the output coupling and is measured by independent reflectivity measurements and shows values around $\eta_{top} = 85 \pm 5 \%$.
% \blue{Rajouter ici une discussion sur l'out coupling qui doit forcément être mesuré pour passer de la brillance à la probabilité d'occupation}.
For QD1, the maximum observed brightness is $21\%$ and corresponds to a $85 \pm 1 \%$ occupation probability.

These experiments have been repeated on two other devices, QD2 and QD3. In these cases, a remarkably high polarized brightness has been measured: $B_p = 28 \pm 5\%$ for QD2 and $B_p = 33 \pm 5 \%$ for QD3. They both show a similarly high hole occupation probability with $\langle P_h \rangle = 85 \pm 1\%$ for QD2 and with $ \langle P_h \rangle = 91 \pm 1 \%$ for QD3.

\section{Performances operating as single-photon sources.}
\label{sec_source}

\begin{figure}
	\centering
	\includegraphics[width=7cm]{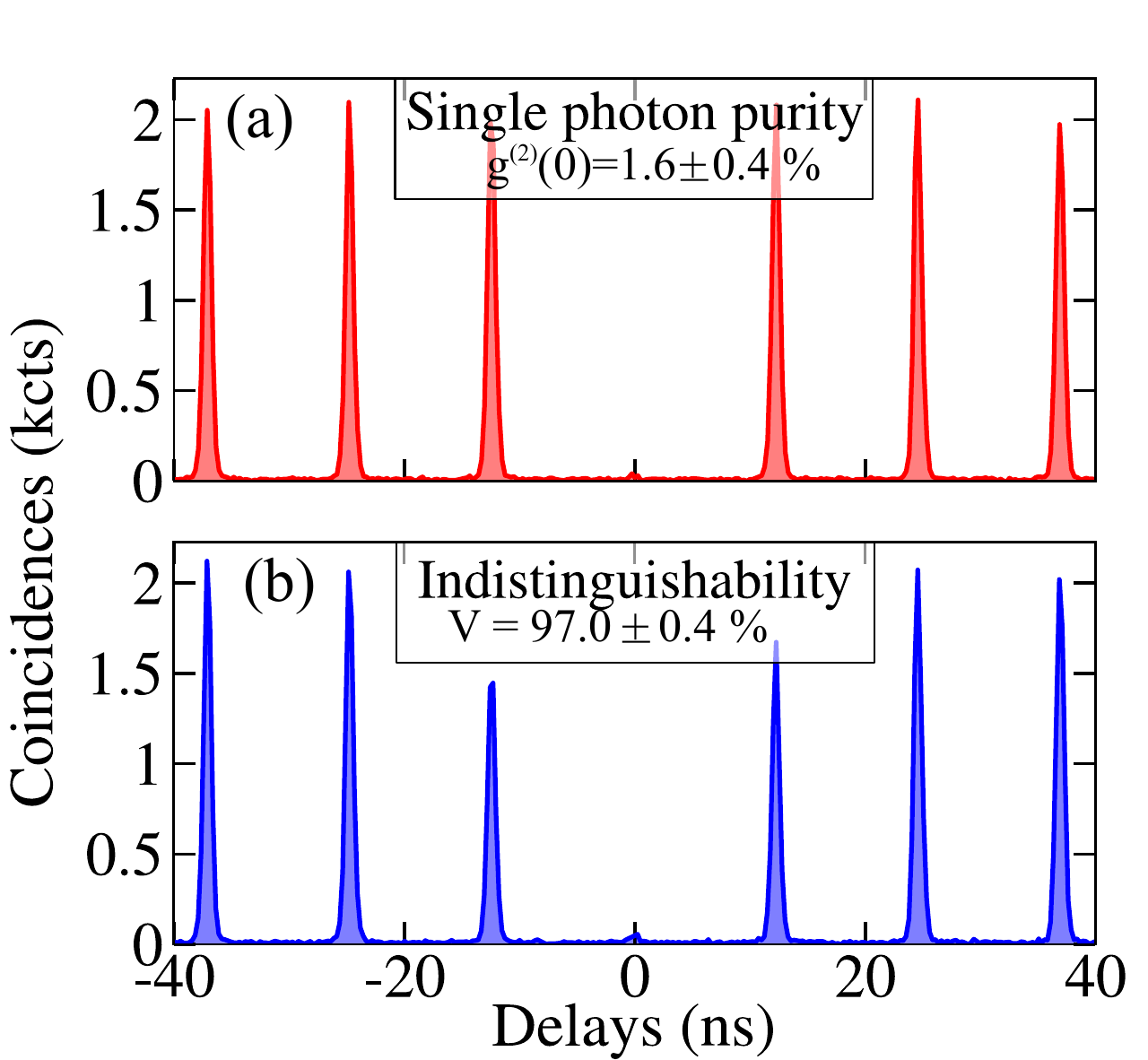}
	\caption{Quantum performances of the cQED devices. Single photon purity and photon indistinguishability are estimated by HBT (a) and HOM (b) experiments for device QD1 (with a Fabry-Pérot etalon to better suppress the reflected laser).}.
	\label{fig_plot_g2_hom}
\end{figure}

Finally, the quantum properties of the cQED devices are investigated by assessing performances as single photon sources.
The single-photon purity is evaluated by the short timescale zero-delay intensity correlations, measuring the $g^{(2)}(0)$ with a Hanbury-Brown and Twiss (HBT) experiment~\cite{Brown1958}.
% with a Hanbury-Brown and Twiss experiment~\cite{Brown1958} as previously, but we are now interested in the zero-delay behavior, which contains information about the single-photon purity of the resonance fluorescence.
A spectral filter ($30\pico\meter$ bandwidth) is inserted in the collection setup to further suppress the spectrally-wide excitation laser and phonon side band emission. The results obtained on QD1 show a good single-photon purity $g^{(2)}(0) = 1.6 \pm 0.4\%$ as displayed in Fig.~\ref{fig_plot_g2_hom}(a).

The single-photon indistinguishability is evaluated by coalescence measurements via the Hong-Ou-Mandel (HOM) effect~\cite{Hong1987}.
% , in which two consecutively emitted photons interfere on a beam splitter and coalesce on the two outputs as much as they are indistinguishable.
This is performed experimentally using a path-unbalanced Mach-Zender interferometer~\cite{Loredo2019} where the difference of delay between the two arms is set to be equal to the laser period $T_R$. Therefore, two single-photons generated by two immediately separated laser pulses can interfere.
Fig.~\ref{fig_plot_g2_hom}(b) displays the experimental results obtained on QD1: the measured photon indistinguishability is  at the state of the art with a raw HOM visibility  $V = 97 \pm 0.4\%$.

% Fig.~\ref{fig_plot_g2_hom}(b) displays the experimental results obtained on QD1: the measured photon indistinguishability is also high, with an indistinguishability  $V = 97 \pm 0.4\%$. These results ensure the good quantum properties of the spin-photon interaction in the interfaces mentionned in this paper.

\section{Conclusion}

We have provided a set of experimental and technological tools for the controlled realization of singly-charged QD-photon interfaces. The key points of our work are: facilitating the hole confinement with a tunneling barrier; using in-plane magnetic-field spectroscopy to identify the trion transition under non-resonant excitation; defining the cavities with the proper geometry to match the targeted trion transition using the in-situ lithography technique; and optically injecting a single hole with a quasi-resonant pumping scheme. The resulting charged QD-photon interfaces have then been used to monitor in real time the jumping of the hole in and out of the quantum dot on a time scale of tens of microseconds. Autocorrelation measurements give access to the hole tunnelling time, and to the hole occupation probability $\langle P_h \rangle$ which has been shown to reach large values, between $85\%$ and $91\%$, for multiple devices.  The high-quality of the spin-photon interfacing devices was evidenced by state-of-the-art performances of the devices operated  as bright sources of pure and indistinguishable photons.
% These performances should allow our devices to behave as efficient photon receivers in the future, potentially permiting the realization of deterministic photonic quantum gates \cite{Hu2008, Hu2009, Bonato2010, Rosenblum2011}.

\section*{Acknowledgments}

This work was partially supported by the ERC PoC PHOW,
the French Agence Nationale pour la Recherche (grant ANR SPIQE and ASTRID LIGHT),
the QuantERA network (project HIPHOP),
the French RENATECH network and a public grant overseen by the French National
Research Agency (ANR) as part of the Investissements d'Avenir programme (Labex
NanoSaclay, reference ANR-10-LABX-0035). J.C.L. and C.A. acknowledge support from
Marie Sk\l{}odowska-Curie Individual Fellowships SMUPHOS and SQUAPH, respectively.

\bibliographystyle{apsrev4-1}
\bibliography{BiblioThesis}

\end{document}